\documentclass [12pt]{article}

\usepackage     {amsthm}
\usepackage     {graphicx}
\usepackage     {mathrsfs}
\usepackage     {amsfonts}
\usepackage     {makeidx}         
\usepackage     {multicol}        

\usepackage     {graphicx}
\usepackage     {mathrsfs}
\usepackage     {amsfonts}
\usepackage     {makeidx}         
\usepackage     {multicol}        

\title{\bf Complementarity problems for electro-neutral charged bodies}
\author{A.A. Kolpakov (Universit\'{e} de Fribourg, Suisse) \\ A.G. Kolpakov (Marie Curie Fellow, Novosibirsk, Russia) }
\date{}
\begin{document}

\maketitle

\noindent{\textit{Solutions to the complementarity problem constructed in} \textrm{[1]}, \textit{generally, possess non-zero total charge. In natural sciences, bodies possessing non-zero total charge (ions and similar object) are considered as specific objects. Bodies possessing zero total charge (electro-neutral bodies) are considered as general case objects. This paper presents a solution to the complementarity problem for electro-neutral bodies. The solution is constructed under the condition that the volumes of the bodies are small.}

\medskip \medskip\medskip
\noindent
{\bf The statement of the complementarity problem for charged bodies} 
\medskip

\noindent
Let us consider two bodies with axisymmetric distributions of charges  $\phi({\bf x})$  and $\psi({\bf x})$  as shown in Fig.1. In this case, the interaction of two bodies is measured by the interaction force in the direction $Ox_3$ [1]. We restrict the translational degrees of freedom in $Ox_1$- and $Ox_2$-directions and allow a translation along the $Ox_3$-axis only. 

\begin{figure}[h]
\centering
\includegraphics[scale=0.55]{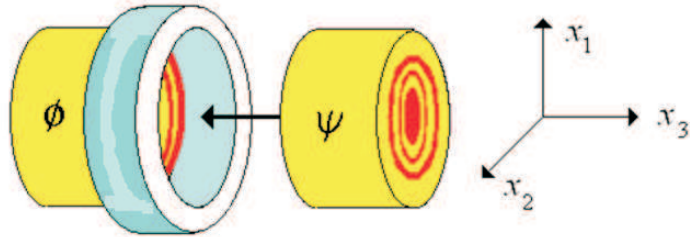}
\caption{\it A pair of bodies with axisymmetric distributions of charges}  
\label{Fig1}
\end{figure} 

The total interaction force $I\left\langle \phi, \psi \right\rangle$  for the given pair of bodies with distributions of charges $\phi({\bf x})$ and $\psi({\bf x})$ over them in the ``one touches another'' position is [1]
\begin{eqnarray}   &  \displaystyle
I\left\langle \phi, \psi \right\rangle 
= \int_Q  \int_Q   R(|{\bf x}-{\bf y}|) \phi({\bf x}) d{\bf x} \psi({\bf y}) d{\bf y}
\label{(1)}
\end{eqnarray}

Let us introduce the operator [1]
\begin{eqnarray}   &  \displaystyle
{\mathbf R} \mbox{ : }\phi({\bf y}) \in L_2(Q) \to  \int_Q   R(|{\bf x}-{\bf y}|) \phi({\bf y}) d{\bf y} \in L_2(Q).
\label{(2)}
\end{eqnarray}

With this notation, we have the following formula for the pair interaction force
 \begin{eqnarray}   &  \displaystyle
I\left\langle \phi, \psi \right\rangle  = ({\mathbf R} \phi, \psi),
\label{(3)}
\end{eqnarray}
where $(\cdot ,\cdot )$ means the standard scalar product in $L_2(Q)$ [2].

The kernel $R(|{\bf x}-{\bf y}|)$  of the integral operator in (2) is the fundamental solution to the electrostatic problem and is singular [3]. We make a simplification and assume that the kernel is a smooth function (thus we assume that the charges belonging to one body do not touch the charges belonging to another body).
     
A body is called electro-neutral if its total charge equals zero. In other words, the distribution $\phi({\bf x})$ has zero average value over the domain $Q$
$$
\left\langle \phi \right\rangle = \int_Q   \phi({\bf x}) d{\bf x}=0
$$
(here one should not mix the averaging operator $ \left\langle \cdot \right\rangle $ with the coupling operation $ \left\langle \mbox{ , } \right\rangle $).

The subspace  $L_0=\{\phi \in L_2(Q) \mbox{ : }  \left\langle \phi, \psi \right\rangle =0 \}$ is the space of charges of electro-neutral bodies.

Generally speaking, the operator ${\mathbf R}$ (2) acts from  $L_0$ to $L_2(Q)$, but not to $L_0$ [1]. In other words, operator ${\mathbf R}$  transforms an electro-neurtal distribution into a distribution, that is not necessarily electo-neutral.

\medskip
\noindent
{\bf Definition}. {\it We say the distributions of charges $\phi({\bf x})$ and $\psi({\bf x})$ to be complementary if  $I\left\langle \phi, \psi \right\rangle>0$ ($\phi({\bf x})$  and $\psi({\bf x})$ attract one another) and to be \textbf{not} complementary if  $I\left\langle \phi, \psi \right\rangle>0$ ($\phi({\bf x})$  and $\psi({\bf x})$ push away one another)}. 

\medskip
The problem is formulated as follows: ``Do there exist four distributions 

\medskip \medskip 
\begin{center}
\begin{tabular}{|l|r|} 
\hline
$\phi$  & $\Phi$
\\ 
\hline
$\psi$  & $\Psi$ 
\\ \hline
\end{tabular}
\end{center}

\medskip \medskip 
\noindent
which belong to $L_0$  and possess the following properties: $\phi$ is complementary to  $\Phi$ and not complementary to  $\phi$, $\psi$   and  $\Psi$;  $\psi$ is complementary to $\Psi$ and not complementary to $\phi$, $\psi$   and  $\Phi$, neither $\Phi$  is complementary to $\Psi$?''

This means that $\phi$  attracts  $\Phi$ and pushes away the others and vice versa,  $\psi$ attracts  $\Psi$ and pushes away the others and vice versa. In all other possible pairs, the distributions push away one another. It is seen that the complementarity problem is a version of the ``lock and key'' problem [4] under the additional condition that the lock ``attracts'' a suitable key and ``pushes away'' all the others. Due to (3), the conditions above may be written in the form

\begin{eqnarray}  &
I\left\langle \phi, \Phi \right\rangle > 0, \label{(4)} \\&
I\left\langle \phi, \phi \right\rangle < 0,
I\left\langle \phi, \psi \right\rangle < 0,
I\left\langle \phi, \Psi \right\rangle < 0,
I\left\langle \Phi, \Phi \right\rangle < 0,
I\left\langle \Phi, \Psi \right\rangle < 0,
\nonumber \\ &
I\left\langle \psi, \Psi \right\rangle > 0, \nonumber \\ &
I\left\langle \psi, \psi \right\rangle < 0,
I\left\langle \psi, \phi \right\rangle < 0,
I\left\langle \psi, \Phi \right\rangle < 0,
I\left\langle \Psi, \Psi \right\rangle < 0,
I\left\langle \Psi, \Phi \right\rangle < 0.
\nonumber
\end{eqnarray} 
Due to the symmetry of the interaction forces with respect to the distribution of charges (see formula (1)), only ten inequalities in (4) are independent. The system of inequalities (4) must be solved under the condition  $\phi, \psi, \Phi, \Psi \in L_0.$

  \medskip \medskip\medskip
\noindent
{\bf Solution to the problem} 
\medskip

\noindent
We have already noticed that $L_0$  is not a proper subspace of the operator ${\mathbf R} $ (2).  As a result, we cannot apply the method developed in [1] to solve the complementarity problem for electro-neutral bodies.
      
We note that $L_0$   is a subspace of  $L_2(Q)$. Thus, $L_2(Q)$ can be represented as the sum:  $L_2(Q)=L_0 \otimes L_0^\bot$ of $L_0$  and its orthogonal complement  $L_0^\bot$ [2]. In the case under consideration,  $L_0^\bot=\{\phi \in L_2(Q) \mbox{ : }  \phi({\bf x})=const \}.$ We denote by  $\mathbf{Pr} $ the operator of projection [2] acting from $L_2(Q)$ to  $L_0$. Then 
\begin{eqnarray}   &  \displaystyle
\mathbf{Pr}\, \phi= \phi({\bf x}) - \left\langle  \phi \right\rangle.
\label{(5)}
\end{eqnarray}     

The linear operator  ${\mathbf R} \mbox{ : } L_2(Q)\to L_2(Q)$, which is compact and self-adjoint [1], has a system of eigenfunctions $\{ \varphi_i \}_{i=1}^\infty$, which form an orthogonal basis in  $L_2(Q)$. In particular, 
\begin{eqnarray}   &  \displaystyle
{\mathbf R} \varphi_i= \lambda_i \varphi_i    
\label{(5)} \\ &  \displaystyle
(\varphi_i, \varphi_j)=\delta_{ij},
\nonumber
\end{eqnarray}
where $\lambda_i$  means the eigenvalue corresponding to the eigenfunction  $\varphi_i$.

We consider two functions  $\mathbf{Pr} \phi_i= \phi_i - \left\langle  \phi_i \right\rangle$ and  $\mathbf{Pr} \phi_j= \phi_j - \left\langle  \phi_j \right\rangle$ and compute the  interaction force for this pair
\begin{eqnarray}   &  \displaystyle
I\left\langle  {\mathbf R} \phi_i, {\mathbf R} \phi_j \right\rangle=
({\mathbf R} \mathbf{ Pr} \varphi_i, \mathbf{ Pr} \varphi_j).    
\label{(7)} 
\end{eqnarray}

By using (5) and (6), we have 
\begin{eqnarray}   &  \displaystyle \label{(8)} 
({\mathbf R} \mathbf{ Pr} \varphi_i, \mathbf{ Pr} \varphi_j)=
\nonumber  \\  &  \displaystyle 
({\mathbf R} (\phi_i - \left\langle  \phi_i \right\rangle), \phi_j - \left\langle  \phi_j \right\rangle)=
 \\  &  \displaystyle 
  =(\lambda_i\phi_i - \left\langle  \phi_i \right\rangle {\mathbf R} 1, \phi_j - \left\langle  \phi_j\right\rangle)=
  \nonumber   \\  &  \displaystyle 
={\lambda_i} (\phi_i, \phi_j)-\left\langle  \phi_i \right\rangle 
   \left\langle  \phi_j, {\mathbf R} 1\right\rangle) 
   -{\lambda_i} \left\langle  \phi_i \right\rangle \left\langle  \phi_j \right\rangle
  +\left\langle  \phi_i \right\rangle \left\langle  {\mathbf R} 1 \right\rangle \left\langle  \phi_j \right\rangle= 
\nonumber  \\  &  \displaystyle 
  {\lambda_i} (\delta_{ij}- \left\langle  \phi_i \right\rangle \left\langle  \phi_j \right\rangle)-
  \left\langle  \phi_i \right\rangle 
  (   \left\langle  \phi_j, {\mathbf R} 1\right\rangle 
   - \left\langle  {\mathbf R} 1 \right\rangle \left\langle  \phi_j \right\rangle).    \nonumber
   \end{eqnarray}  
We use here the equality $(f,C)=C \cdot \left\langle f \right\rangle $, where $C$ is a constant.

With regard to the Schwartz inequality $ \displaystyle \left| \int_Q f({\bf x})g({\bf x})d{\bf x}\right|   \leq ||f|| \cdot ||g||$  [2], we have the following estimates for the terms forming the r.h.p. of (8): 
 
  \begin{eqnarray}   &  \displaystyle \label{(9)} 
| \left\langle  \phi_i \right\rangle|=   \left| \int_Q \phi_i({\bf x})1d{\bf x} \right|   \leq ||\phi_i|| \cdot ||1||=\sqrt{mes Q}, 
 \\[8pt]  &  \displaystyle 
|\left\langle  \phi_j, {\mathbf R} 1\right\rangle| \leq
||\phi_j|| \cdot ||{\mathbf R} 1||=||{\mathbf R} 1||.
  \nonumber
   \end{eqnarray}  

In (9), we use the fact that $||\phi_i||=1$  (where $||\mbox{ }||=1$ means the standard norm in $L_2(Q)$ [2]). Then the terms in the r.h.p. of (8) can be estimated as follows:
 \begin{eqnarray}   &  \displaystyle \label{(10)} 
  \left |{\lambda_i} (\delta_{ij}- \left\langle  \phi_i \right\rangle \left\langle  \phi_j \right\rangle)-
  \left\langle  \phi_i \right \rangle 
  (   \left\langle  \phi_j {\mathbf R} 1\right\rangle 
   - \left\langle  {\mathbf R} 1 \right\rangle \left\langle  \phi_j \right\rangle) \right | \leq
   \\[8pt]  &  \displaystyle 
  \leq mesQ  +\sqrt{mes Q} ( ||{\mathbf R} 1 || + |{\mathbf R} 1 | \sqrt{mes Q}).    \nonumber
   \end{eqnarray} 
   
By definition,
\begin{eqnarray}   &  \displaystyle \label{(11)} 
{\mathbf R} 1= \int_Q   R(|{\bf x}-{\bf y}|) d{\bf y}.    
\end{eqnarray}  
   
We consider a set of domains  $Q(r)$ depending on the real  parameter $r$  and embedded one into another in such a way that  $Q(r_1) \supseteq Q(r_2)$ if  $r_1>r_2$. Then  $\displaystyle 
{\mathbf R} 1(r) = \int_{Q(r)}   R(|{\bf x}-{\bf y}|) d{\bf y}$, being considered as a function of $r$, decreases. Then  $\displaystyle {\mathbf R} 1(r) \leq \int_{Q(r_0)}   R(|{\bf x}-{\bf y}|) d{\bf y}=C_0$ for any $r \leq r_0$  and the r.h.p. of (10)  is not greater than
\begin{eqnarray}   &  \displaystyle \label{(12)} 
mes\,Q  +\sqrt{mes\, Q} \cdot C_0( \sqrt{mes\, Q} + mes\, Q \sqrt{mes\, Q}) \leq C \cdot mes \,Q       \end{eqnarray} 
for any $r \leq r_0$, where  $C< \infty$. Thus, we can write (8) as 
\begin{eqnarray}   &  \displaystyle \label{(13)} 
({\mathbf R} \mathbf{ Pr} \varphi_i, \mathbf{ Pr} \varphi_j)= \lambda_i \delta_{ij} + F_{ij},
\end{eqnarray}
where  $|F_{ij}| \leq C \cdot mes \,Q  $, see (12). 
In order to construct four complementary functions, we consider the functions displayed in Table 1.

\medskip \medskip 
\begin{center}
Table 1. {\it Two pairs of electro-neutral distributions}
\medskip 

\begin{tabular}{|l|r|} 
\hline
	$\phi=\mathbf{ Pr} \varphi_i +\alpha \mathbf{ Pr} \varphi_k$  
& $\Phi = -\mathbf{ Pr} \varphi_i +\alpha \mathbf{ Pr} \varphi_k $
\\ 
\hline
$	\psi=\mathbf{ Pr} \varphi_j +\alpha \mathbf{ Pr} \varphi_k$  
& $\Psi = - \mathbf{ Pr} \varphi_j +\alpha \mathbf{ Pr} \varphi_k $     \\ \hline
\end{tabular}
\end{center}

\medskip \medskip 
\noindent
where  $\varphi_i$, $\varphi_j$ and $\varphi_k$  are eigenfunctions of the operator  ${\mathbf R}$,  $i \neq j$,  $i \neq k$,  $j \neq k$. The functions  $\phi$,  $\Phi$,  $\psi$ and $\Psi$  determined by Table 1 belong to  $L_0$ (all these distributions are electro-neutral, since all of them are sums of projections of the corresponding functions to  $L_0$).  
The interactions corresponding to the functions displayed in Table 1 are the following:
   
I. The interaction of $\phi$ with $\Phi$
   \begin{eqnarray}   &  \displaystyle \label{(14)} 
  ({\mathbf R} \phi, \Phi)=
-({\mathbf R} (\mathbf{ Pr} \varphi_i +\alpha \mathbf{ Pr} \varphi_k), -\mathbf{ Pr} \varphi_i +\alpha \mathbf{ Pr} \varphi_k) =
\\ &  \displaystyle 
= ({\mathbf R} \mathbf{ Pr} \varphi_i , \mathbf{ Pr} \varphi_i )
- \alpha ({\mathbf R} \mathbf{ Pr} \varphi_k, \mathbf{ Pr} \varphi_i )+
\nonumber \\ &  \displaystyle 
+ \alpha ({\mathbf R} \mathbf{ Pr} \varphi_i, \mathbf{ Pr} \varphi_k)
+ \alpha^2 ({\mathbf R} \mathbf{ Pr} \varphi_k,  \mathbf{ Pr} \varphi_k)=
\nonumber \\ &  \displaystyle 
= - \lambda_i -F_{ii} - \alpha (F_{ki}-F_{ik})+ \alpha^2 (\lambda_k + F_{kk})=
\nonumber \\ &  \displaystyle 
=- \lambda_i + \alpha^2 \lambda_k +[-F_{ii} - \alpha (F_{ki}-F_{ik})+ \alpha^2 F_{kk}].
\nonumber 
   \end{eqnarray}

II. The interaction of $\psi$ with $\Psi$
   \begin{eqnarray}   &  \displaystyle \label{(15)} 
  ({\mathbf R} \psi, \Psi)=
({\mathbf R} (\mathbf{ Pr} \varphi_j +\alpha \mathbf{ Pr} \varphi_k), -\mathbf{ Pr} \varphi_j +\alpha \mathbf{ Pr} \varphi_k) =
\\ &  \displaystyle 
=- ({\mathbf R} \mathbf{ Pr} \varphi_j , \mathbf{ Pr} \varphi_j )
- \alpha ({\mathbf R} \mathbf{ Pr} \varphi_k, \mathbf{ Pr} \varphi_j )+
\nonumber \\ &  \displaystyle 
+ \alpha ({\mathbf R} \mathbf{ Pr} \varphi_j, \mathbf{ Pr} \varphi_k)
+ \alpha^2 ({\mathbf R} \mathbf{ Pr} \varphi_k,  \mathbf{ Pr} \varphi_k)=
\nonumber \\ &  \displaystyle 
= - \lambda_j -F_{jj} - \alpha (F_{kj}-F_{jk})+ \alpha^2 (\lambda_k + F_{kk})=
\nonumber \\ &  \displaystyle 
=- \lambda_j + \alpha^2 \lambda_k +[-F_{jj} - \alpha (F_{ki}-F_{ik})+ \alpha^2 F_{kk}].
\nonumber 
   \end{eqnarray}
   
III. The interaction of $\phi$ with $\psi$
   \begin{eqnarray}   &  \displaystyle \label{(16)} 
  ({\mathbf R} \phi, \psi)=
({\mathbf R} (\mathbf{ Pr} \varphi_i +\alpha \mathbf{ Pr} \varphi_k), \mathbf{ Pr} \varphi_j +\alpha \mathbf{ Pr} \varphi_k) =
\\ &  \displaystyle 
= ({\mathbf R} \mathbf{ Pr} \varphi_i , \mathbf{ Pr} \varphi_j )
+ \alpha ({\mathbf R} \mathbf{ Pr} \varphi_k, \mathbf{ Pr} \varphi_j )+
\nonumber \\ &  \displaystyle 
+ \alpha ({\mathbf R} \mathbf{ Pr} \varphi_i, \mathbf{ Pr} \varphi_k)
+ \alpha^2 ({\mathbf R} \mathbf{ Pr} \varphi_k,  \mathbf{ Pr} \varphi_k)=
\nonumber \\ &  \displaystyle 
= F_{ij} - \alpha (F_{kj}+F_{ik})+ \alpha^2 (\lambda_k + F_{kk})=
\nonumber \\ &  \displaystyle 
= \alpha^2 \lambda_k +[F_{ij} - \alpha (F_{kj}+F_{ik})+ \alpha^2 F_{kk}].
\nonumber 
   \end{eqnarray} 
   
IV. The interaction of $\phi$ with $\Psi$
   \begin{eqnarray}   &  \displaystyle \label{(17)} 
  ({\mathbf R} \phi, \Psi)=
({\mathbf R} (\mathbf{ Pr} \varphi_i +\alpha \mathbf{ Pr} \varphi_k), -\mathbf{ Pr} \varphi_j +\alpha \mathbf{ Pr} \varphi_k) =
\\ &  \displaystyle 
= ({\mathbf R} \mathbf{ Pr} \varphi_i , \mathbf{ Pr} \varphi_j )
- \alpha ({\mathbf R} \mathbf{ Pr} \varphi_k, \mathbf{ Pr} \varphi_j )+
\nonumber \\ &  \displaystyle 
+ \alpha ({\mathbf R} \mathbf{ Pr} \varphi_i, \mathbf{ Pr} \varphi_k)
+ \alpha^2 ({\mathbf R} \mathbf{ Pr} \varphi_k,  \mathbf{ Pr} \varphi_k)=
\nonumber \\ &  \displaystyle 
= F_{ij} - \alpha (F_{kj}-F_{ik})+ \alpha^2 (\lambda_k + F_{kk})=
\nonumber \\ &  \displaystyle 
= \alpha^2 \lambda_k +[F_{ij} - \alpha (F_{kj}-F_{ik})+ \alpha^2 F_{kk}].
\nonumber 
   \end{eqnarray}   
 
V. The interaction of $\Phi$ with $\Psi$
   \begin{eqnarray}   &  \displaystyle \label{(18)} 
  ({\mathbf R} \Phi, \Psi)=
({\mathbf R} (\mathbf{ Pr} \varphi_i +\alpha \mathbf{ Pr} \varphi_k), -\mathbf{ Pr} \varphi_j +\alpha \mathbf{ Pr} \varphi_k) =
\\ &  \displaystyle 
= -({\mathbf R} \mathbf{ Pr} \varphi_i , \mathbf{ Pr} \varphi_j )
- \alpha ({\mathbf R} \mathbf{ Pr} \varphi_k, \mathbf{ Pr} \varphi_j )+
\nonumber \\ &  \displaystyle 
+ \alpha ({\mathbf R} \mathbf{ Pr} \varphi_i, \mathbf{ Pr} \varphi_k)
+ \alpha^2 ({\mathbf R} \mathbf{ Pr} \varphi_k,  \mathbf{ Pr} \varphi_k)=
\nonumber \\ &  \displaystyle 
= -F_{ij} - \alpha (F_{kj}-F_{ik})+ \alpha^2 (\lambda_k + F_{kk})=
\nonumber \\ &  \displaystyle 
= \alpha^2 \lambda_k +[F_{ij} - \alpha (F_{kj}-F_{ik})+ \alpha^2 F_{kk}].
\nonumber 
   \end{eqnarray}

VI. The interaction of $\psi$ with $\Phi$
   \begin{eqnarray}   &  \displaystyle \label{(19)} 
  ({\mathbf R} \psi, \Phi)=
({\mathbf R} (\mathbf{ Pr} \varphi_j +\alpha \mathbf{ Pr} \varphi_k), -\mathbf{ Pr} \varphi_i +\alpha \mathbf{ Pr} \varphi_k) =
\\ &  \displaystyle 
= -({\mathbf R} \mathbf{ Pr} \varphi_j , \mathbf{ Pr} \varphi_i )
- \alpha ({\mathbf R} \mathbf{ Pr} \varphi_k, \mathbf{ Pr} \varphi_i )+
\nonumber \\ &  \displaystyle 
+ \alpha ({\mathbf R} \mathbf{ Pr} \varphi_j, \mathbf{ Pr} \varphi_k)
+ \alpha^2 ({\mathbf R} \mathbf{ Pr} \varphi_k,  \mathbf{ Pr} \varphi_k)=
\nonumber \\ &  \displaystyle 
= F_{ji} - \alpha (F_{ki}-F_{jk})+ \alpha^2 (\lambda_k + F_{kk})=
\nonumber \\ &  \displaystyle 
= \alpha^2 \lambda_k +[F_{ij} - \alpha (F_{ki}-F_{jk})+ \alpha^2 F_{kk}].
\nonumber 
   \end{eqnarray}

VII. The interaction of $\phi$ with $\phi$
   \begin{eqnarray}   &  \displaystyle \label{(20)} 
  ({\mathbf R} \phi, \phi)=
({\mathbf R} (\mathbf{ Pr} \varphi_i +\alpha \mathbf{ Pr} \varphi_k), \mathbf{ Pr} \varphi_i +\alpha \mathbf{ Pr} \varphi_k) =
\\ &  \displaystyle 
= -({\mathbf R} \mathbf{ Pr} \varphi_i , \mathbf{ Pr} \varphi_i )
- \alpha ({\mathbf R} \mathbf{ Pr} \varphi_k, \mathbf{ Pr} \varphi_i )+
\nonumber \\ &  \displaystyle 
+ \alpha ({\mathbf R} \mathbf{ Pr} \varphi_i, \mathbf{ Pr} \varphi_k)
+ \alpha^2 ({\mathbf R} \mathbf{ Pr} \varphi_k,  \mathbf{ Pr} \varphi_k)=
\nonumber \\ &  \displaystyle 
= \lambda_i + F_{ii} - \alpha (F_{ki}-F_{ik})+ \alpha^2 (\lambda_k + F_{kk})=
\nonumber \\ &  \displaystyle 
= \lambda_i + \alpha^2 \lambda_k +[F_{ii} - \alpha (F_{ki}-F_{ik})+ \alpha^2 F_{kk}].
\nonumber 
   \end{eqnarray}

VIII. The interaction of $\psi$ with $\psi$
   \begin{eqnarray}   &  \displaystyle \label{(21)} 
  ({\mathbf R} \psi, \psi)=
({\mathbf R} (\mathbf{ Pr} \varphi_j +\alpha \mathbf{ Pr} \varphi_k), \mathbf{ Pr} \varphi_j +\alpha \mathbf{ Pr} \varphi_k) =
\\ &  \displaystyle 
= -({\mathbf R} \mathbf{ Pr} \varphi_j , \mathbf{ Pr} \varphi_j )
- \alpha ({\mathbf R} \mathbf{ Pr} \varphi_k, \mathbf{ Pr} \varphi_j )+
\nonumber \\ &  \displaystyle 
+ \alpha ({\mathbf R} \mathbf{ Pr} \varphi_j, \mathbf{ Pr} \varphi_k)
+ \alpha^2 ({\mathbf R} \mathbf{ Pr} \varphi_k,  \mathbf{ Pr} \varphi_k)=
\nonumber \\ &  \displaystyle 
= \lambda_j + F_{jj} - \alpha (F_{kj}+F_{jk})+ \alpha^2 (\lambda_k + F_{kk})=
\nonumber \\ &  \displaystyle 
= \lambda_j + \alpha^2 \lambda_k +[F_{jj} - \alpha (F_{kj}+F_{jk})+ \alpha^2 F_{kk}].
\nonumber 
   \end{eqnarray}

IX. The interaction of $\Phi$ with $\Phi$
   \begin{eqnarray}   &  \displaystyle \label{(22)} 
  ({\mathbf R} \Phi, \Phi)=
({\mathbf R} (-\mathbf{ Pr} \varphi_i +\alpha \mathbf{ Pr} \varphi_k), -\mathbf{ Pr} \varphi_i +\alpha \mathbf{ Pr} \varphi_k) =
\\ &  \displaystyle 
= ({\mathbf R} \mathbf{ Pr} \varphi_i , \mathbf{ Pr} \varphi_i )
- \alpha ({\mathbf R} \mathbf{ Pr} \varphi_k, \mathbf{ Pr} \varphi_i )-
\nonumber \\ &  \displaystyle 
- \alpha ({\mathbf R} \mathbf{ Pr} \varphi_i, \mathbf{ Pr} \varphi_k)
+ \alpha^2 ({\mathbf R} \mathbf{ Pr} \varphi_k,  \mathbf{ Pr} \varphi_k)=
\nonumber \\ &  \displaystyle 
= \lambda_i + F_{ii} - \alpha (F_{ki}+F_{ik})+ \alpha^2 (\lambda_k + F_{kk})=
\nonumber \\ &  \displaystyle 
= \lambda_i + \alpha^2 \lambda_k +[F_{ii} - \alpha (F_{ki}+F_{ik})+ \alpha^2 F_{kk}].
\nonumber 
   \end{eqnarray}  
   
X. The interaction of $\Psi$ with $\Psi$
   \begin{eqnarray}   &  \displaystyle \label{(23)} 
  ({\mathbf R} \Psi, \Psi)=
({\mathbf R} (-\mathbf{ Pr} \varphi_j +\alpha \mathbf{ Pr} \varphi_k), -\mathbf{ Pr} \varphi_j +\alpha \mathbf{ Pr} \varphi_k) =
\\ &  \displaystyle 
= ({\mathbf R} \mathbf{ Pr} \varphi_j , \mathbf{ Pr} \varphi_j )
- \alpha ({\mathbf R} \mathbf{ Pr} \varphi_k, \mathbf{ Pr} \varphi_j )-
\nonumber \\ &  \displaystyle 
- \alpha ({\mathbf R} \mathbf{ Pr} \varphi_j, \mathbf{ Pr} \varphi_k)
+ \alpha^2 ({\mathbf R} \mathbf{ Pr} \varphi_k,  \mathbf{ Pr} \varphi_k)=
\nonumber \\ &  \displaystyle 
= \lambda_j + F_{jj} - \alpha (F_{kj}+F_{jk})+ \alpha^2 (\lambda_k + F_{kk})=
\nonumber \\ &  \displaystyle 
= \lambda_i + \alpha^2 \lambda_k +[F_{jj} - \alpha (F_{kj}+F_{jk})+ \alpha^2 F_{kk}].
\nonumber 
   \end{eqnarray}

Since the operator ${\mathbf R}$  is negatively determined [1], all of its eigenvalues   $\{ \lambda_i \}_{i=1}^\infty$ are negative. 

The r.h.p.'s of formulas (14)-(23) involve the following independent parameters: the real number  $\alpha$ and certain combinations of the quantities $\{ F_{ij} \}$  with various indices (see the expressions in square brackets). The absolute value of the terms in the square brackets in (14)-(18) is not greater than
 \begin{eqnarray}   &  \displaystyle \label{(24)} 
 F+2\alpha F+\alpha^2F \leq F(1+\alpha)^2  mes\, Q , 
   \end{eqnarray} 
where  $F=max\{F_{mn};m,n=i,j,k \}$. For  $\alpha <1$, the r.h.p. of (24) is not greater than  $4Fmes\, Q$. 
If we take
 \begin{eqnarray}   &  \displaystyle \label{(25)} 
 \alpha < \min \left( \sqrt{\frac{\lambda_i}{2\lambda_k}}, \sqrt{\frac{2\lambda_i}{\lambda_k}}\right)  , 
   \end{eqnarray} 
and
\begin{eqnarray}   &  \displaystyle \label{(26)} 
 mes\, Q < \frac{1}{2} \min \left( \sqrt{\frac{\lambda_i}{2F}}, \sqrt{\frac{2\lambda_i}{2F}}\right)  , 
\end{eqnarray} 
then the r.h.p.'s of (14) and (15) (the interactions of $\phi$ with $\Phi$ and of $\psi$ with  $\Psi$) are positive and the r.h.p.'s of formulas (16)-(23) (i.e. all the other interactions of four distributions of charges displayed in Table 1) are negative, thus (4) is satisfied. 

This means that there exist two pairs of electro-neutral distributions of charges: one pair is $\phi$ and  $\Phi$, another pair is $\psi$ and $\Psi$ and all the other possible pairs of distributions are not complementary.
     
Inequality (26) means that the volume of the bodies is bounded from above. Note that although the volume  $mes\, Q$ is small, it is a finite positive number.

\medskip  \medskip \medskip
\noindent
{\bf Acknowledgements}. A.G.K. was supported through Marie Curie actions FP7, project PIIF2-GA-2008-219690.

\medskip \medskip \medskip
\noindent
\noindent{\bf References}

\noindent
1.	A.A. Kolpakov, A.G. Kolpakov, Complementarity problems for two pairs of charged bodies, arXiv:1205.5157 

\noindent
2.	K. Yosida, Functional Analysis. Springer, Berlin; 1996.

\noindent
3.	J.D. Jackson, Classical Electrodynamics. 3rd Ed Wiley, NY; 1998.

\noindent
4.	J. Stenesh, Biochemistry. Plenum Press, NY; 1998.

\end{document}